\documentclass[]{spie}  

\usepackage{amsmath,amsfonts,amssymb}
\usepackage{graphicx}
\usepackage{braket}
\usepackage[colorlinks=true, allcolors=blue]{hyperref}

\title{Towards a Better Understanding of OPD Limitations for Higher Sensitivity and Contrast at the VLTI.
}
\author[a,e]{B. Courtney-Barrer} 
\author[b]{J. Woillez} 
\author[c]{R. Laugier}
\author[c]{A. Bigioli}
\author[a]{N. Schuhler}
\author[a]{P. Guajardo}
\author[a]{V. Lizana}
\author[a]{N. Behara}
\author[d]{F. Eisenhauer}
\author[e]{M. Ireland}
\author[a]{X. Haubois}
\author[c]{D. Defrère}
\affil[a]{European Southern Observatory, Alonso de Cordova 3107, Vitacura, Santiago, Chile}
\affil[b]{European Southern Observatory Headquarters, Karl-Schwarzschild-Straße 2, 85748 Garching, Germany}
\affil[c]{Institute of Astronomy, KU Leuven, Celestijnenlaan 200D, 3001 Leuven, Belgium}
\affil[d]{Max Planck Institute for extraterrestrial Physics, Giessenbach-straße 1, 85748 Garching, Germany}
\affil[e]{Research School of Astronomy \& Astrophysics, Australian National University, ACT 2611, Australia}

\authorinfo{Further author information: (Send correspondence to Benjamin Courtney-Barrer)\\: E-mail: bcourtne@eso.org}

\begin{document} 
\maketitle

\begin{abstract}
Precise control of the optical path differences (OPD) in the Very Large Telescope Interferometer (VLTI) was critical for the characterization of the black hole at the center of our Galaxy - leading to the 2020 Nobel prize in physics. There is now significant effort to push these OPD limits even further, in-particular achieving 100nm OPD RMS on the 8m unit telescopes (UT's) to allow higher contrast and sensitivity at the VLTI. This work calculated the theoretical atmospheric OPD limit of the VLTI as 5nm and 15nm RMS, with current levels around 200nm and 100nm RMS for the UT and 1.8m auxillary telescopes (AT's) respectively, when using bright targets in good atmospheric conditions. We find experimental evidence for the $f^{-17/3}$ power law theoretically predicted from the effect of telescope filtering in the case of the ATs which is not currently observed for the UT's. Fitting a series of vibrating mirrors modelled as dampened harmonic oscillators, we were able to model the UT OPD PSD of the gravity fringe tracker to $<1nm/\sqrt{Hz}$ RMSE up to 100Hz, which could adequately explain a hidden $f^{-17/3}$ power law on the UTs. Vibration frequencies in the range of 60-90Hz and also 40-50Hz were found to generally dominate the closed loop OPD residuals of Gravity. Cross correlating accelerometer with Gravity data, it was found that strong contributions in the 40-50Hz range are coming from the M1-M3 mirrors, while a significant portion of power from the 60-100Hz contributions are likely coming from between the M4-M10. From the vibrating mirror model it was shown that achieving sub 100nm OPD RMS for particular baselines (that have OPD$\sim$200nm RMS) required removing nearly all vibration sources below 100Hz. 
\end{abstract}


\section{Introduction}
\label{sec:VLTI}
\begin{figure} [h]
   \begin{center}
   \begin{tabular}{c} 
   \includegraphics[height=8cm]{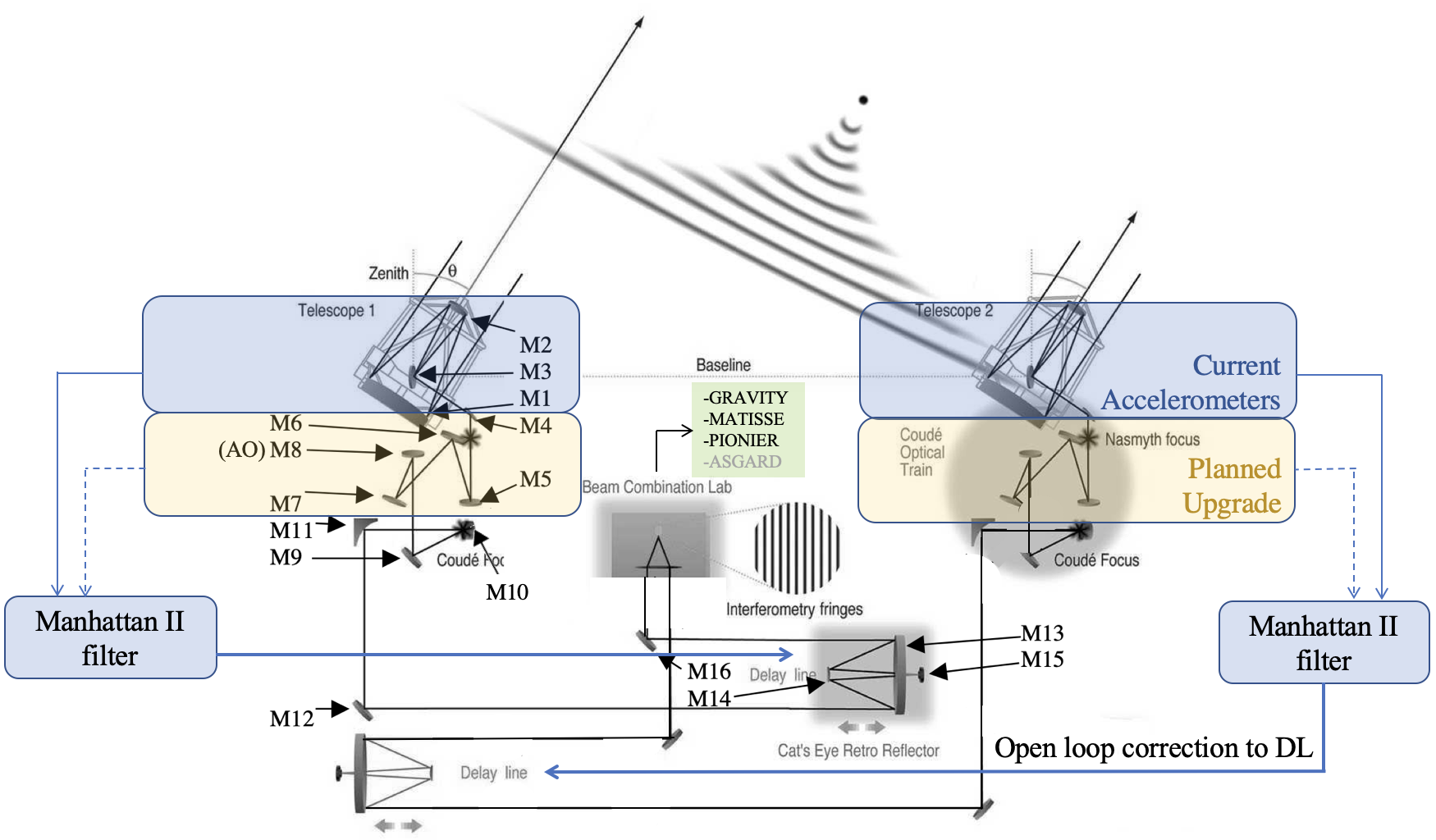}
   \end{tabular}
   \end{center}
   \caption  
   { \label{fig:VLTI optical train}
   The VLTI optical train indicating where the current accelerometers are and where the new ones will be installed for the Manhattan II upgrade. This image was adapted from ref. \citenum{guisard_2003}}
   \end{figure}

With the push for higher sensitivity and contrast at the VLTI comes a key requirement for improved control of the optical path differences (OPD) between interfering beams. A major goal for new generation instruments is to get OPD levels as measured on the Gravity fringe tracker with the 8m Unit Telescopes (UT) below 100nm RMS – with current levels around 200nm RMS under good atmospheric conditions\cite{grav_1st_light_2017,defrere_hi5_2018}. This will unlock a range of outstanding progress in various fields of near infrared interferometry including characterising photospheres from known (and potentially unknown) exoplanets, studying young stellar objects, binarity, exozodiacal disks and AGN's\cite{defrere_hi5_2018, gravity+_white_paper}. The current VLTI facility holds three instruments; Pionier which covers H-Band\cite{pionier_2011}, Matisse covering LMN-bands\cite{matisse_2022} and Gravity\cite{grav_1st_light_2017} covering K band which also includes the Gravity fringe tracker to co-phase either Gravity or Matisse in the new GRA4MAT mode of operation\cite{matisse_user_manual}. Currently new upgrades are underway with the planned Asgard visitor suite\cite{defrere_hi5_2018} and Gravity+ which will allow for wide-field off-axis fringe tracking, improved sensitivity, and laser guide star adaptive optics, with the aim of giving access to targets as faint as K = 22 mag \cite{gravity+_white_paper}. \\
Understanding specifically what is limiting OPD levels is difficult due to the huge number of instrument dependent and independent subsystems that are coupled to the VLTI optical path.
This optical path is shown in figure \ref{fig:VLTI optical train}. There are roughly 20 mirrors to get each beam to a VLTI instrument from the UT's, each mechanically coupled to the surrounding environment which impart vibrations onto the lights OPD. To initially compensate this a series of accelerometers were installed on the UTs - 4 on the primary mirror (M1), 1 on the secondary mirror (M2) and 2 on the tertiary mirror (M3). The signals from these are passed through a filter and then fed forward to apply an open loop correction to the delay lines \cite{lieto_2007_MN2_orig}. This system is known collectively as Manhattan II (MN2). With the campaign to reduce VLTI's OPD there is a planned extension of the MN2 system with new accelerometers being installed on the M4-M8 and a potential upgrade to the MN2 filter as indicated in figure \ref{fig:VLTI optical train}. For this work we seek to understand the fundamental atmospheric limits of the VLTI, identify frequencies and OPD contributions of dominant vibration sources, and briefly touch on the impacts of cross talk from AO residuals to piston in the case of fiber coupled instruments.

\section{Theoretical OPD Limits in the Atmosphere}
\label{sec:atm_theory}  
The spatial power spectral density (PSD) of a wavefronts phase passing through earths atmosphere is typically modelled by the Von Karman spectra, which in the limit of the outerscale $(L_0)$ going to infinity gives the Kolmogorov model \cite{conan_2000}.
\begin{equation}\label{eq:von-karman-spectra}
W_{\phi}(k) = 0.0029r_0^{-5/3} \left(  k^2+\left(\frac{1}{L_0}\right)^2 \right)^{-11/6}
\end{equation}
\\\\
Re-iterating primary results of Conan\cite{conan_1995}: For some quantity G that is related to the wavefront phase $\phi(r)$ via a convolution with a spatial filtering function $M_G(r)$ (e.g. a telescope pupil), the spatial power spectral density (PSD) of G is given as: 
\begin{equation}
    W_G(\textbf{k}) = |\hat{M_G}(\textbf{k})|^2 W_{\phi}(\textbf{k})
\end{equation}
where we denote the Fourier Transform $\mathcal{F}[M_G] = \hat{M_G}$, $W_\phi(k)$ is the spatial PSD of the wavefront phase $\phi(r)$ and k is the spatial frequency vector. Making the Taylor frozen turbulence approximation for some wind velocity \textbf{V} blowing along the x-axis, this quantity can then be turned into the temporal PSD: 
\begin{equation}\label{eq:fund_temporal_eq}
    w_G(\nu) = \frac{1}{V}\int^{\infty}_{-\infty}\left|\hat{M_G}\left(\frac{\nu}{V},f_y\right)\right|^2 W_{\phi}\left(\frac{\nu}{V},f_y\right)df_y
\end{equation}
where $\nu$ is the temporal frequency (Hz), $f_y$ the spatial frequency along the y-axis and V the wind speed along the x-axis. 
\\\\
Basic results from Conan\cite{conan_1995} for the temporal PSD of atmospheric piston over a disk pupil are: 
\begin{equation} \label{eq:PSD_p}
    w_p(\nu) = \frac{4}{V}\int^{\infty}_{-\infty}\left| \frac{J_1(\pi D q)}{\pi D q} \right|^2 W_{\phi}(q) df_y
\end{equation}
and for the differential piston (OPD) over some baseline B:
\begin{equation} \label{eq:PSD_dp}
    w_{dp}(\nu) = \frac{4}{V}\left(1-\sin^2{\left(\frac{\pi B \nu}{V} \right) } \right) \int^{\infty}_{-\infty}\left| \frac{J_1(\pi D q)}{\pi D q} \right|^2 W_{\phi}(q) df_y
\end{equation}
Where $q=\sqrt{(\nu/V)^2+f_y^2}$, and $J_1$ is the first order Bessel function. \\\\
Since analytic solutions to equations \ref{eq:PSD_p} \& \ref{eq:PSD_dp} are cumbersome and non-existent in some cases, it is important to understand the physics of what is happening and map this to the asymptotic limits of these equations. If we first consider a infinitesimally small point ($\hat M_G=1$) and imagine a phase screen passing over it. The change in phase (piston) between two moments close in time is purely determined from the statistics of the atmosphere, which is modelled by the Von-Karman spectra. Integration of \ref{eq:PSD_p} in this case leads to a natural -8/3 power law in the pistons temporal PSD (which is also seen in the differential piston). Moving to a finite area and considering the average phase over this area, which is our definition of piston, we can see that the change in piston between two moments in time is critically determined by how far the phase screen has passed over the area in consideration. For small time increments (high temporal frequencies), most of the phase within the area is the same as in the previous moment, only changing values at the areas perimeter. Therefore the change in piston is small - much smaller then the case of a singular point, and is not only determined by the statistics of the atmosphere, but also by area and velocity of the phase screen. This is known as the telescope filtering effect, which in the case of a circular area (telescope pupil) leads to a -17/3 power law in the temporal PSD. However if larger time increments are considered (low temporal frequencies) such that the phase screen over the area from one time step to the next is completely new - once again we return to the case that the change in piston is only determined by statistics of the atmosphere, and equation \ref{eq:PSD_p} returns again to a -8/3 power law in the temporal PSD. The transition knee between this high frequency and low frequency power law naturally occurs at the inverse time it takes for a phase screen to pass completely over the telescope, reported in literature to occur at $\sim0.3V/D$ where V is the wind velocity of the turbulence and D is the telescope diameter. In the case of differential piston (OPD) there is also a very low frequency regime determined by how long it takes for the phase screen to pass over the entire baseline (B) which to first order follows a -2/3 power law relation with a knee transition at $\sim V/B$. In the case that the wind is orthogonal to the baseline the differential piston PSD is simply twice that of the piston.

\subsection{Impact of waveguide coupling} \label{sec:atm_theory_waveguide}
\begin{figure*}[h]
    \centering
    \includegraphics[width=10cm]{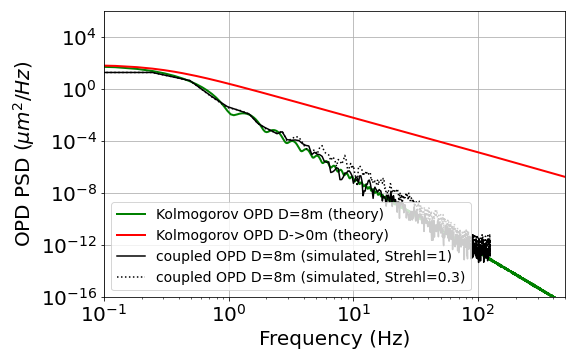}
    \caption{Temporal PSD of the theoretical Kolmogorov OPD for a non-waveguide coupled system considering an 8m telescope diameter (green) and also the limit as the diameter goes to zero (red). These are calculated at a wavelength = 2.2um, seeing = 0.8", coherence time = 3ms, and outerscale = 25m over a 44m baseline. This is compared to simulations of the OPD after coupling to the Gravity waveguide (black curves) for a 8m telescope in the case of a perfect wavefront (K Strehl=1) and AO corrected wavefront (K Strehl=0.3).}
    \label{fig:atm_limits_theory}
\end{figure*}

In the case of single mode waveguide coupling, as is used for Gravity, the situation becomes more complex as the spatial distribution of the input fields phasors gets mapped to a single waveguide phasor. The piston in this case is defined as the angle of this phasor which is determined by the overlap of the waveguides mode and the input electric field filtered by the telescope pupil \cite{roddier_1988_fiber_coupling}. This piston can no longer be represented simply as a convolution of the phase with a spatial filtering function, and hence the treatment of Conan\cite{conan_1995} cannot be used. We consider the scalar product between two quantities X and Y defined over a spatial coordinate $\textbf{r}$ with weight W such that: 
\begin{equation}
    \braket{X|Y}_W = \int\int_{R^2} W(\textbf{r})X(\textbf{r})Y(\textbf{r})d^2\textbf{r}
\end{equation}
Given an input field into a telescope pupil $P$:
\begin{equation}
    E(\textbf{r}) = E_0 \exp{( i \phi(\textbf{r}) )}
\end{equation}
and complex waveguide mode $M(\textbf{r})$ represented in the pupil plane, the general expression for the normalized phasor ($\Omega_\phi$) coupled to a single waveguide mode M from a telescope pupil P is:
\begin{equation}
    \Omega = \frac{ \braket{E|M}_{P} }{ (\braket{E|E}_{P} \braket{M|M}_{P})^{1/2} }
\end{equation}
Where the amplitude of the coupled field is given by $|\Omega|E_0$ and the phase is given by the argument of $\Omega$ respectively. Hence the waveguide coupled piston (p) is defined:
\begin{equation} \label{eq:coupled_piston}
    p = \mathrm{arg}\ \Omega 
\end{equation}
There is no general analytic solution for equation \ref{eq:coupled_piston} therefore simulations were performed using the AOtools python package \cite{aotools_2019} to generate rolling Kolmogorov phase screens (with finite outerscales). Waveguide coupling was calculated using Roddier's \cite{roddier_1988_fiber_coupling} Gaussian approximation of the LP01 mode of a circular single mode fiber, which in the focal plane is given as:
\begin{equation}
    {E}(\rho) = \frac{c_H}{\omega} exp{\left( -\frac{\rho^2}{\omega^2}\right)} 
\end{equation}
Where $\rho=\sqrt{(k_x^2 + k_y^2)}$ is the image plane radial coordinate, $\omega=a(0.65+1.619/V^{1.5} +2.879/V^6)$, $c_H=\sqrt{\left( \frac{2n_{cl}}{\pi}\right)}\left( \frac{\epsilon_o}{\mu}\right)^{1/4}$, $n_{cl}$ is the fiber cladding refractive index, $\epsilon_o$ and $\mu$ are the permittivity/permeability of space, while a is the fiber core radius. 
We consider waveguide parameters representative of the Gravity waveguide as discussed in \cite{lacour_2014} and simulate the waveguide coupled OPD both in the case of a perfect wavefront (Kband Strehl=1) with just the piston moving up and down, and also with partial AO correction - correcting the first 30 Zernike modes to simulate an effective Kband Strehl ratio of 0.3. This is representative of MACAO performance for Kmag$<$14. Both the perfect wavefront and AO corrected wavefronts are simulated from the same rolling phase screens with finite outerscale equal to the measured Paranal mean\cite{conan_2000}. In the case of the perfect wavefront, just the piston term was extracted from the original phase screens. The waveguide coupled phase was unwrapped to avoid $\pi$ discontinuities which (incorrectly) adds a $f^{-2}$ power law into the PSD. We also consider the temporal PSD of the Kolmogorov OPD in the limit of a singular point (D$\rightarrow$0m). These theoretical and simulated OPD's for the waveguide and non-waveguide coupled cases are shown in figure \ref{fig:atm_limits_theory}. The waveguide coupled OPD follows the same behaviour as the non-coupled Kolmogorov OPD, however in the case of a non-perfect wavefront we see increased OPD at higher frequencies. This has negligible ($\sim$1\%) impact on the total OPD in open loop situations since lower frequencies dominate the OPD, however in closed loop operations this difference can become more significant. Future work will provide further characterization of this effect. 

\section{Comparison of Gravity's OPD residuals to the Atmospheric Limit  }

\subsection{The Gravity Fringe Tracker}\label{sec:grav_fringe_tracker}
The Gravity fringe tracker is the core OPD control system for co-phasing the Gravity instrument to allow long integration times on a science target, achieving typical OPD RMS of 200nm in good atmospheric conditions.  It is now also being used to co-phase other instruments such as Matisse  and potentially the Asgard visitor instrument suite \cite{defrere_hi5_2018} in the future. A more complete overview of the Gravity fringe tracker can be found in \citenum{lacour_2019} while the theory and simulations of the fringe trackers Kalman filter can be found in reference \citenum{menu_2012_fringe_tracker_kalman}. The Gravity fringe tracker operates in parallel with the Gravity science detectors which all operate in K-band. There are various modes of fringe tracking which are constantly evolving, Gravity wide \cite{gravity_wide_2022} being the most recently available mode. In general Gravity can be operated in on-axis or off-axis as well as combined or split polarization modes with 3 available integration times of 0.85ms, 3ms and 10ms (see the Gravity user manual for more details). The Fringe tracker consists of 3 core control systems; Group delay control, phase delay Kalman control, and a Feed forward predictor based on the action of the actuator. The full control block diagram can be found in figure 2 in \citenum{lacour_2019}. The output sends commands to actuate Gravity's internal delay lines, with large offsets sent to the main VLTI delay lines. The OPD residuals and the fringe trackers pseudo open loop can be calculated from linear combinations of the measured and Kalman estimated OPD, delay line positions and modulation phase. This is mathematically described in \citenum{lacour_2019} and explicitly detailed in the Gravity Pipeline User Manual in the context of Gravity data products \cite{gravity_user_manual}. In-addition the pseudo open loop of the Gravity fringe tracker can be calculated by undoing the fringe tracker commands on the OPD residuals. This process is also described in references \citenum{lacour_2019,gravity_user_manual}. The transfer function of the Gravity fringe tracker can then be estimated by taking the ratio of the cross spectral density of the OPD residuals to pseudo open loop OPD for a given baseline. Examples of the Gravity transfer function can be found in reference \citenum{grav_1st_light_2017}.

\subsection{The Atmospheric Limit of the VLTI} \label{sec:atm_limit_vlti}
\begin{figure*}[h]
    \centering
    \includegraphics[width=14cm]{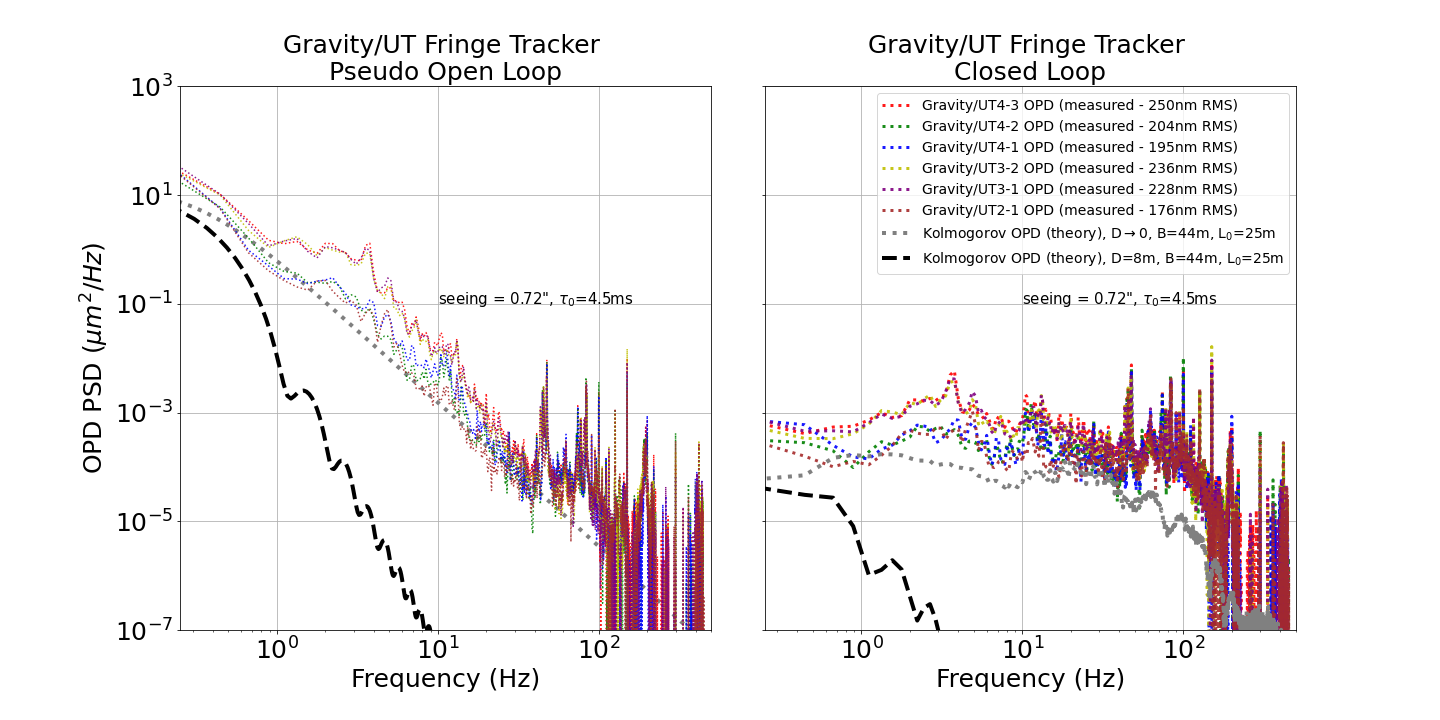}
    \includegraphics[width=14cm]{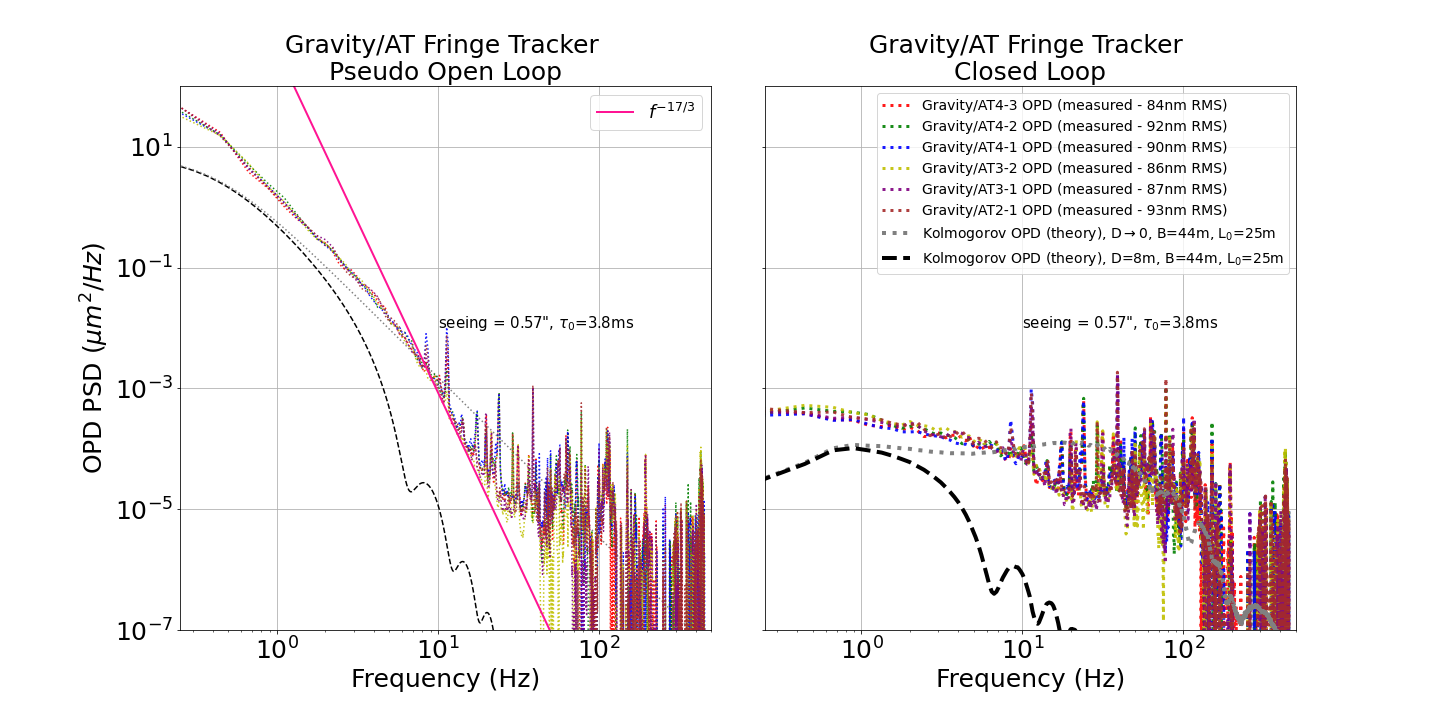}
    \caption{Pseudo open (left column) and closed loop (right column) OPD measurements from the Gravity fringe tracker on the UT (top row) and AT (bottom row) telescopes taken for bright calibrator stars in near median atmospheric conditions at Paranal. These are compared to the theoretical Kolmogorov OPD with/without applying the baseline mean transfer function of the Gravity fringe tracker using either the correct telescope diameter (black, dashed line), or the limit of a point aperture (grey, dots). }
    \label{fig:atm_limit_vlti}
\end{figure*}

We calculate the atmospheric limit of the VLTI for both the UT and AT telescopes by applying transfer function of the Gravity fringe tracker to the theoretical Kolmogorov OPD as discussed in section \ref{sec:atm_theory}. The transfer function is calculated as discussed in section \ref{sec:grav_fringe_tracker} using bright calibrator stars (Kmag = 0.9 \& 1.8 for the UT and AT's respectively), the fastest fringe tracking possible ($\sim$0.85ms) with on-axis, split polarisation modes in near median atmospheric conditions (seeing = 0.5"-0.75", $\tau_0$ = 3-5ms). Theoretical calculations were made using the mean of these atmospheric conditions over the observations. The OPD RMS is calculated by integrating the respective OPD PSD between 0.1 - 500Hz. We compare these theoretical OPD limits to the OPD residuals measured by the Gravity fringe tracker for the respective calibrator stars. Table \ref{tab:opd_closed_loop} outlines these results while figure \ref{fig:atm_limit_vlti} compares the measured OPD PSDs in pseudo open and closed loop to the atmospheric limits. 
\begin{table}[h!]
    \centering
    \begin{tabular}{|c|c|c|}
        \hline
        Setup & measured & atmospheric limit \\
              &     OPD RMS  & OPD RMS  \\
        \hline\hline
        UT GRAVITY & 214$\pm$25nm & 5$\pm$0.6nm \\
        \hline
        AT GRAVITY & 89$\pm$4nm & 15$\pm$0.6nm \\
        \hline
    \end{tabular}
    \caption{Measured RMS OPD residuals (baseline mean $\pm$ standard deviation) on the Gravity fringe tracker with bright targets vs Kolmogorov atmospheric OPD RMS after applying the Gravity fringe tracker's transfer function for the UT and AT telescopes. Measured and theoretical results were done in near median atmospheric conditions as indicated in figure \ref{fig:atm_limit_vlti} with finite outscale $L_0=25m$ as reported by \cite{conan_2000}. The OPD RMS is calculated by integrating the respective OPD PSD between 0.1 - 500Hz.}
    \label{tab:opd_closed_loop}
\end{table}
The measured OPD RMS on the UTs is found to be 42 times greater then the theoretical limit, while the measured OPD RMS on the AT's it is only 6 times above the theoretical limit for the given conditions. Increasing the outerscale effectively increases the given limits, in the limiting case of an infinite outerscale we calculate the OPD RMS to be increased by 80\% and 13\% of those reported in table \ref{tab:opd_closed_loop} for the UT and ATs respectively. Interestingly, the PSD of the UT telescopes OPD follows more closely the theoretical behaviour of Kolmogorov turbulence for a point aperture, following a near $f^{-8/3}$ power law up to high frequencies, with no clear signs of the $f^{-17/3}$ slope theoretically expected from telescope filtering\cite{conan_1995}. Reasons for this will be discussed in subsequent sections. On the other hand, for the AT's, we do indeed see a clear transition from the $f^{-8/3}$ power law to a $f^{-17/3}$ slope occuring at roughly 8Hz, which is consistent with a wind speed of 50m/s. Such wind speeds are typical at 12km, in the so called jet stream layer, which is typically a dominant layer interms of atmospheric turbulence\cite{cantalloube_2020_jetstream}. This slope can be observed up to around 40Hz which then becomes dominated by vibrations. Despite showing the $f^{-17/3}$ the measured OPD PSD is still higher then the theoretical model in this region which is probably due to the affect of averaging multi-directional winds which was not accounted in the model. To the authors knowledge this is the first published measurement of a stellar interferometer showing a $f^{-17/3}$ power law as theoretically predicted. Many previous models and discussions assumed only a $f^{-8/3}$ power law actually exists for the atmospheric OPD due to lack of experimental evidence otherwise \cite{lacour_2019, menu_2012_fringe_tracker_kalman,colavita_1987}, citing multi-directional wind and integrated Cn2 profiles as the potential cause\cite{menu_2012_fringe_tracker_kalman}. Therefore the observational evidence for the existence of this steep $f^{-17/3}$ power law has important implications for the fundamental limits of the VLTI and other stellar interferometers. For example, assuming the theoretical limits of Kernel nulling, reaching the above reported atmospheric OPD limit would allow contrasts below 10$^{-6}$ to be reached on the VLTI/UT for bright targets\cite{martinache_2018_kernel_nulling}. Furthermore, from the measured and theoretical limits of the pseudo open loop PSD's, we may calculate the respective temporal structure function using relations outlined in reference \citenum{nightingale_1991}. From this we calculate that if the OPD atmospheric limit on the UTs is obtained, it would take roughly 14 times longer for the open loop OPD to reach 1radian RMS then is currently being achieved on the UT's. This would allow significantly longer integration times on the fringe tracker and therefore access to fainter targets.

\section{VLTI/UT Vibrations} \label{sec:vlti_vibrations}
\begin{figure*}[ht]
    \centering
    \includegraphics[width=14cm]{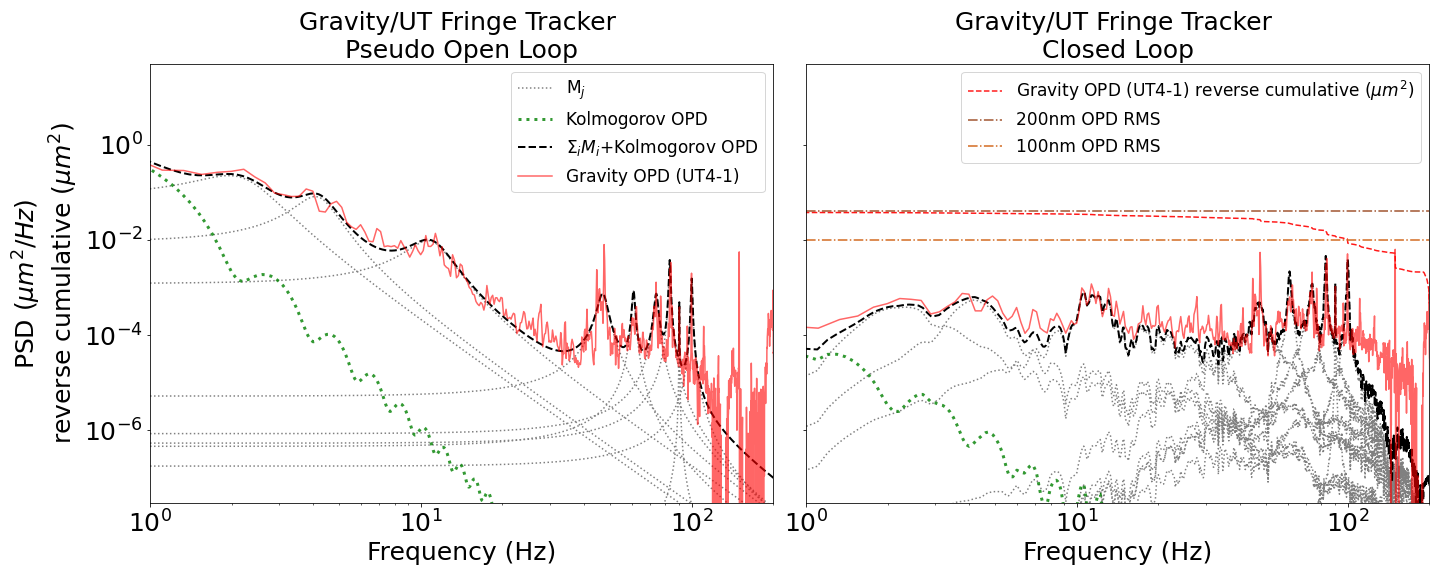}
    \caption{Fitting the OPD PSD of the Gravity fringe tracker (red curve) between 1-100Hz as the superposition (black, dashed curve) of the Kolmogorov atmosphere (green curve) and a (minimal) series dampened harmonic oscillators (grey curves) driven by a normalized impulse response (white spectrum), we achieve an RMSE=0.43nm/$\sqrt{Hz}$. Left and right plots show the results in pseudo open and closed loop operations respectively.}
    \label{fig:harmonic_oscillator_fit_psd}
\end{figure*}

To first order, each mirror in the VLTI optical train can be treated as dampened harmonic oscillator which, depending on the level of dampening, transfers displacements from the surrounding environment with a $f^{-2} – f^{-4}$ frequency roll-off away from its natural resonance frequency \cite{vib_control_textbook_2003}. A critical observation is that this roll-off is in the range of the theoretical $f^{-8/3}$ Kolmogorov power law, but slower than the  $f^{-17/3}$ power law from telescope filtering. Hence an operational environment with of impulse like movements (white seismic noise) can limit the OPD measured beyond the resonance frequency of the system. This is potentially why the effect of telescope filtering is not typically observed in the OPD for large telescope \cite{menu_2012_fringe_tracker_kalman}. Further more it is clear from figure \ref{fig:atm_limit_vlti} that the current measured OPD in Gravity/UT fringe tracker show many broad vibration peaks at low frequencies in comparison to the AT's. We attempt to model the observed OPD PSD in the Gravity/UT fringe tracker between 1-100Hz for the UT4-1 baseline analyzed in section \ref{sec:atm_limit_vlti} as a sum of harmonic oscillators added to theoretical Kolmogorov OPD. We use the minimum possible number of oscillators to reasonably capture the PSD trend between 1-100Hz. We do not attempt to have physically realistic values for the fitted parameters since there is considerable degeneracy, instead we simply assume and impulse response (white spectrum) in the seismic driving force normalized to 1N/$\sqrt{Hz}$ and then fit the natural frequency and dampening terms in the harmonic oscillators to capture the behaviour. The steady state displacement of a dampened harmonic oscillator caused from arbitrary input driving force $F\sin(\omega t)$ at angular frequency $\omega$ can be defined by the transfer function \cite{vib_control_textbook_2003}:
\begin{equation}
    H(\omega) = \frac{1/m}{\omega_0^2-\omega^2+i\Gamma \omega}
\end{equation}
Where m is mass, $\omega_0$ is the oscillators natural frequency, and $\Gamma$ is twice the damping ratio times the natural frequency. Figure \ref{fig:harmonic_oscillator_fit_psd} shows the respective fit to both the pseudo open and closed loop OPD PSD, indicating the individual contributions from each hypothetical mirror (harmonic oscillator) and the atmosphere. Modelling only 9 mirrors (the VLTI track has over 20 mirrors per UT!) we could achieve a RMSE=0.43nm/$\sqrt{Hz}$ between 1-100Hz in closed loop. We then calculate the OPD RMS contribution from each hypothetical mirror after applying the Gravity fringe tracker transfer function to simulate the closed loop residuals. We also calculate the cumulative reduction in the total measured (closed loop) OPD by Gravity from removing contributions from each respective 'mirror', ordered from highest to lowest OPD RMS as shown in table \ref{tab:harmonic_oscillator_table}. The most harmful vibrations are at 83Hz, 74Hz, 47Hz, 61Hz, which collectively contribute 144nm OPD RMS. Attenuating or ultimately removing these would, in this case, reduce Gravity's OPD on the UT4-1 baseline from 194nm potentially down to 129nm RMS. Additionally removing the remaining low frequency ($<15Hz$) contributions and a few narrow band high frequency contributions would ultimately get the PSD below 100nm RMS. The rest of the OPD contributors lay beyond 100Hz as can be seen in the reverse cumulative PSD shown on the right (red, dashed curve) in figure \ref{fig:harmonic_oscillator_fit_psd}. The low Q-factor oscillators fitted to the low frequencies in the PSD can clearly explain why the $f^{-17/3}$ slope is not seen. The origin of these will be discussed further in the next section. In comparison, for the ATs (figure \ref{fig:atm_limit_vlti}) we do not observe any broad (low Q-factor) peaks until beyond at least 40Hz which could explain why the $f^{-17/3}$ slope can be observed.  We note that the particular vibrations detected here is for a single observation and baseline and does not bear statistical significance. A more statistical analysis will be done in subsequent sections. 
\begin{table}[h!]
    \centering
    \begin{tabular}{|c||c|c|c|c|||c|}
        \hline
        $M_j$ & $\omega_0/2\pi$ & $\Gamma$ & 1/m & OPD$_j$ RMS & cumulative residual \\ 
         & [Hz] & [Hz] & [kg$^{-1}$] & [nm] & $\sqrt{OPD_{grav}^2 - \Sigma_j OPD^2_j }$\\
        \hline\hline
        1 & 83 & 2.5 & 200 & 87 & 173nm \\
        \hline
        2 & 74 & 4.0 & 200 & 69 & 159nm \\
        \hline
        3 & 47 & 5.0 & 222 & 66 & 145nm \\
        \hline
        4 & 61 & 3.0 & 100 & 65 & 129nm \\
        \hline
        5 & 11 & 5.0 & 167 & 60 & 114nm \\
        \hline
        6 & 100 & 1.2 & 33 & 36 & 109nm \\
        \hline
        7 & 4.2 & 3.0 & 67 & 35 & 103nm \\
        \hline
        8 & 2.2 & 3.0 & 56 & 29 & 99nm \\
        \hline
        9 & 90 & 1.0 & 12.5 & 17 & 97nm \\
        \hline
    \end{tabular}
    \caption{Fitted parameters for each mirror (dampened harmonic oscillator) used to fit the OPD PSD of the Gravity fringe tracker on the UT4-1 baseline, and the respective closed loop OPD RMS contribution, with mirrors order from highest to lowest OPD contribution. The final column indicates the cumulative reduction in the (closed loop) measured OPD if each OPD contribution from the given mirror (row) is removed (in quadrature). }
    \label{tab:harmonic_oscillator_table}
\end{table}


\subsection{Vibration contributions seen in M1-M3}
\begin{figure*}[h]
    \centering
    \includegraphics[width=13cm]{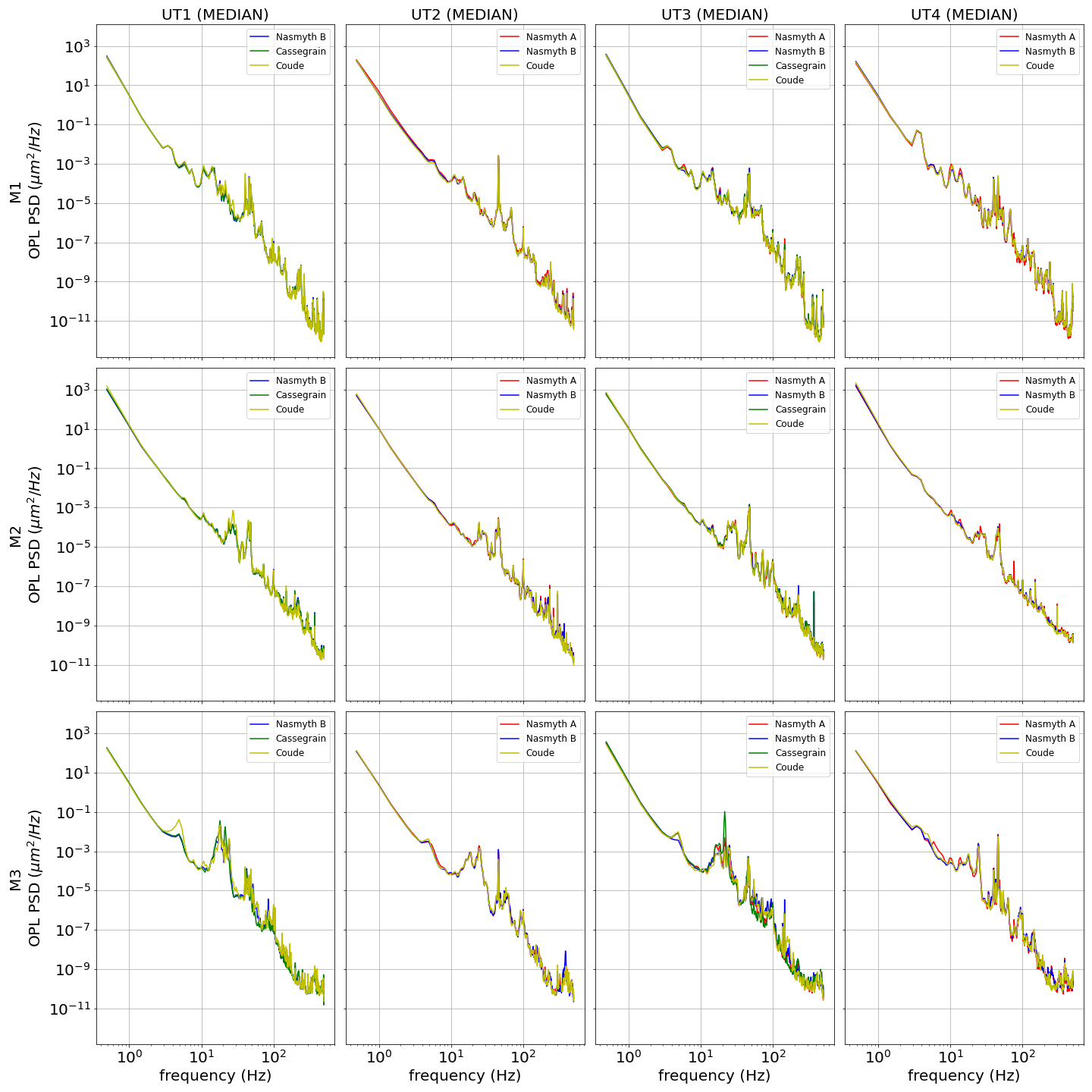}
    \caption{Median PSDs of the OPL calculated for each UT mirror per focus with the current MN2 accelerometer data. }
    \label{fig:mn2_foci_psds}
\end{figure*}

\begin{figure*}[h]
    \centering
    \includegraphics[width=13cm]{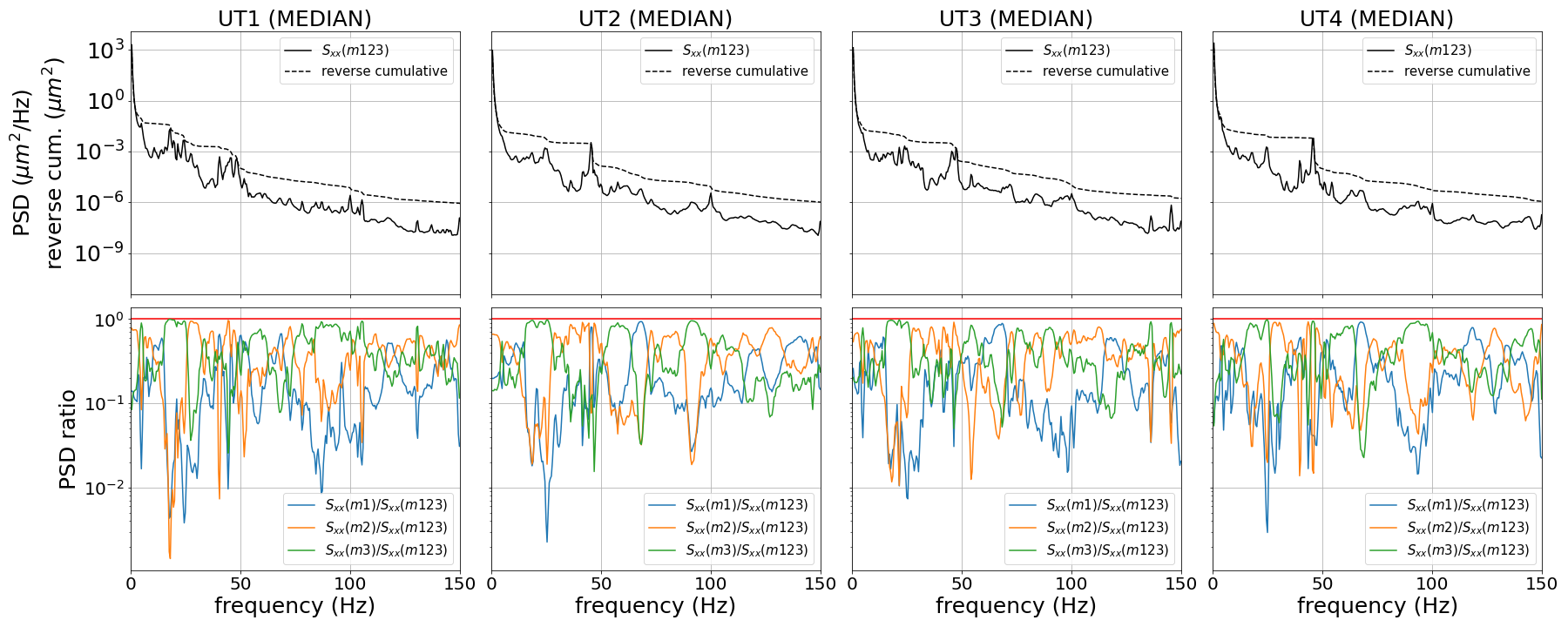}
    \caption{[Top] Median OPL PSD and reverse cumulative PSD of the combined (M123) piston geometry for each UT filtered for Coude focus. [Bottom] The ratio of individual mirrors piston PSD to the combined geometry (M123)}
    \label{fig:mn2_contr}
\end{figure*}

We look at 6 months of MN2 data from Jan-June 2021, considering at random ten second samples (with 1kHz sampling) taken from all current MN2 acelerometers during night operations. These samples are linearly combined to represent the total acceleration in the piston mode for each mirror (M1-M3) and then double integrated to get an OPL. For each piston time series PSD's are calculated and statistical analysis was performed on the PSD's across the entire period using 0.5Hz binning. This consisted of an average of 636 samples per UT per focus while the telescope was in a guiding state. 
No data was available for various foci during this period since:
\begin{itemize}
    \item No instrument was installed on the UT1 Nasmyth-A focus 
    \item VISIR was being installed on the UT2 Cassegrain focus 
    \item ERIS was being installed on the UT4 Cassegrain focus
\end{itemize}
Figure \ref{fig:mn2_foci_psds} shows the median PSDs for each UT mirror with accelerometer data per focus. Quite significant differences can be observed between the same mirrors on different UTs – indicating significantly different vibrational environments. Not surprisingly, for each UT we see the greatest focal dependence in median PSD levels for the M3– albeit not large differences.  Furthermore, the vibration signature on each mirror for a given UT seems to be unique. However we do see a general trend across the UT's for a given mirror in the broad frequency bands that have the most dominant contribution to the mirrors OPL. These are published in \ref{tab:mirror_freq_bands_most_OPL}. The 2, 4, 11Hz low Q-factor vibration peaks attributed to masking the $f^{-17/3}$ when fitting the dampened harmonic oscillators (figure \ref{fig:harmonic_oscillator_fit_psd}) are similarly seen in the median PSD of the M1 and M3 at similar OPL levels. We also see a significant contribution between 15-25Hz in the M3 which is where the current Manhattan II filter is optimized.  
\begin{table}[h!]
    \centering
    \begin{tabular}{|c|c|}
        \hline
        Location & Frequency bands with \\
                 & large OPL contributions \\
        \hline\hline
         M1 & 2-5 Hz \\ 
            & 8-20 Hz \\
            & 40-50 Hz \\
         \hline
         M2 & 20-30Hz \\
            & 30-50Hz \\ 
         \hline
         M3 & 3-8Hz \\
            & 15-25Hz \\
            & 40-60Hz \\ 
         \hline
    \end{tabular}
    \caption{Frequency bands for each mirror that generally have the most dominant contribution to the mirrors OPL for a given UT}
    \label{tab:mirror_freq_bands_most_OPL}
\end{table}
By taking the ratio of individual mirrors piston PSD to the combined geometry (M123) filtered for Coude focus (which is of interest for the optimization of the Manhattan II digital filter), figure \ref{fig:mn2_contr} shows clear regimes where a particular mirror dominates the OPL for each UT in addition to the PSD and reverse cumulative PSD of the combined  (M123) piston geometry. Most notably we see the general trends that low frequency vibrations up to ~15Hz have roughly equal contributions by all mirrors (M1-M3). The M3 typically dominates the vibrations between $\sim$15-30Hz. dominant vibrations in the $\sim$30-50Hz range is highly UT dependent, albeit typically dominated by either the M2 and/or M3. The $\sim$50-70Hz band is typically dominated by the M2. Beyond this the contributions are again UT dependent and less significant in terms of the total OPL. 

\subsection{ Vibration seen on Gravity Fringe Tracker vs Manhattan II}
\begin{figure*}[h]
    \centering
    \includegraphics[width=10cm]{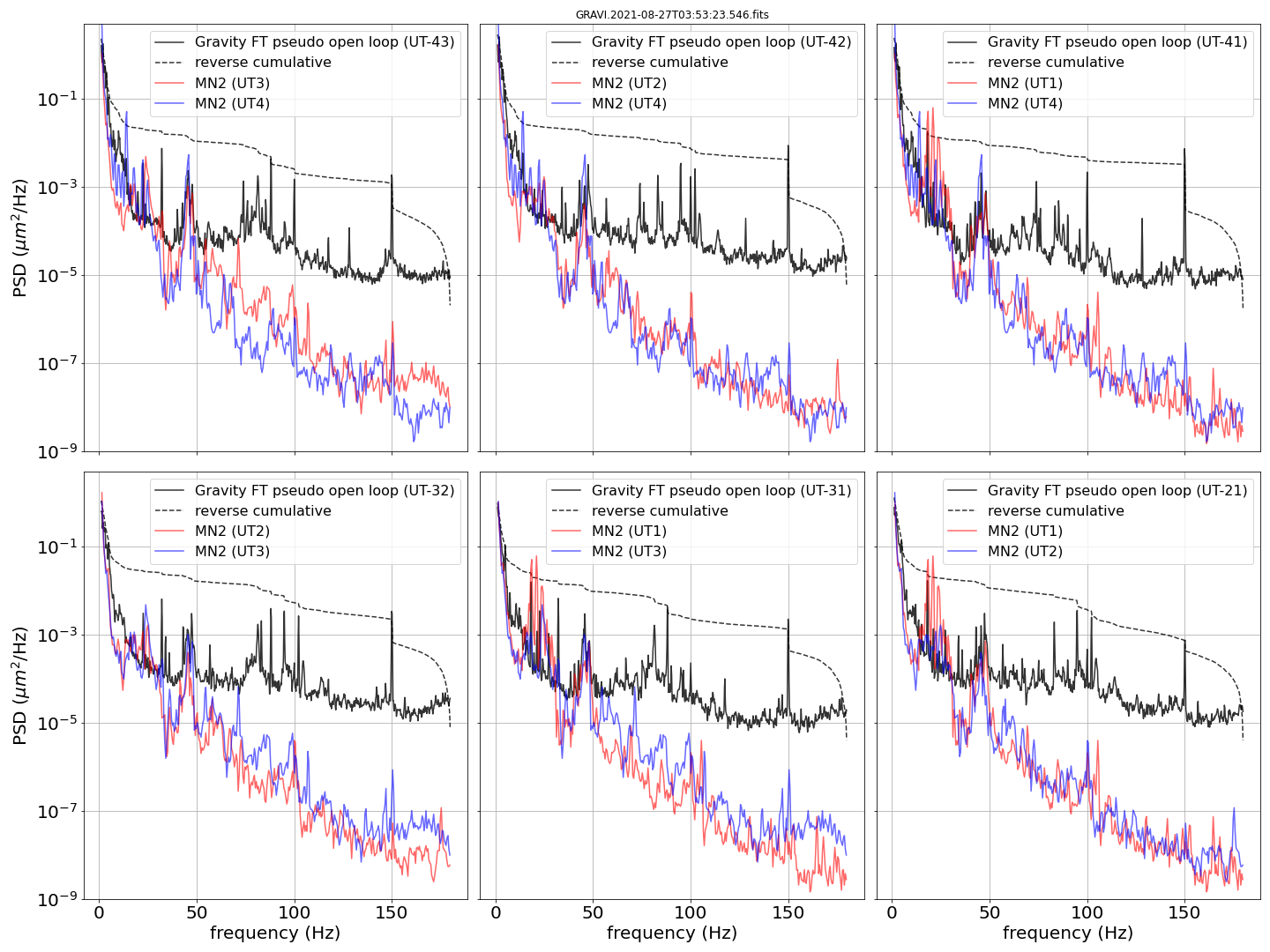}
    \caption{Example PSDs from Aug 2021 comparing the Gravity fringe tracker's OPD in pseudo open loop for each baseline (including the reverse cumulative PSD) vs simulataneous accelerometer (double integrated) measurements of the piston OPL on respective UT's between M1-M3}
    \label{fig:grav_y_mn2_psds}
\end{figure*}

\begin{figure*}[h]
    \centering
    \includegraphics[width=14cm]{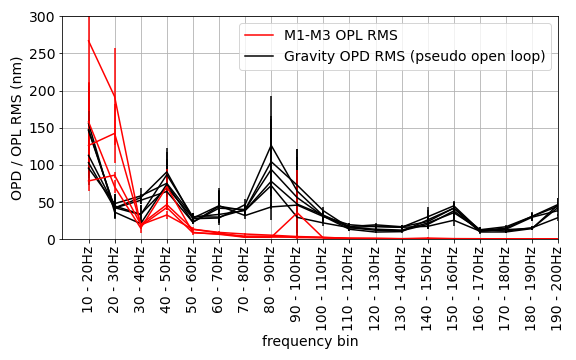}
    \caption{10Hz binned integral of the Gravity fringe tracker's pseudo open loop OPD PSD (black) for each baselines vs the 10Hz binned integral of the measured OPL PSD in the M1-M3 for each UT telescope. each point and error bars indicate the mean and standard deviation across the observations considered for the given baseline or telescope. It is clear that the most damaging domains in Gravity's pseudo open loop are 10-20Hz, 40-50Hz, 80-100Hz.}
    \label{fig:grav_v_mn2_freq_bi}
\end{figure*}

We compare vibrations detected on the Gravity fringe tracker in pseudo open loop to those detected in the current MN2 accelerometers. For this study we randomly select 1 bright target (K$<$7) per month for four recent VLTI/UT runs with MACAO using with the fastest fringe tracker mode ($<$1ms) in good atmospheric conditions (seeing $\sim$ 0.6”). Table \ref{tab:gravity_ft_obs_table} outlines the details of the observations. 
\begin{table}[h!]
    \centering
    \begin{tabular}{|c||c|c|c|}
        \hline
        Timestamp & Kmag & Seeing (") & coherence time (ms) \\
        \hline\hline
        2021-07-25T06:00:18.278 & 5.07 & 0.65 & 4.2 \\
        \hline
        2021-08-27T03:53:23.546 & 5.23 & 0.51 & 5.7 \\
        \hline
        2022-01-23T04:24:58.306 & 4.53 & 0.56 & 12.9 \\
        \hline
        2022-02-20T06:01:51.058 & 6.79 & 0.55 & 13.5 \\
        \hline
    \end{tabular}
    \caption{Details of the observations used to compare vibrations detected on the Gravity fringe tracker in pseudo open loop to those detected in the current MN2 accelerometers.}
    \label{tab:gravity_ft_obs_table}
\end{table}
For each baseline we extracted the raw simultaneous accelerometer data for the corresponding UT's and processed the combined (M1-M3) piston displacement. An example of the combined Gravity and MN2 PSD's are shown in figure \ref{fig:grav_y_mn2_psds}. 
Using the python scipy.signal package \cite{2020SciPy-NMeth} we define vibration peaks as those that exceed the local (25Hz domain) median by a factor of 3. For each detected peak a series a features are calculated including the peak width, prominence, and peak frequency. From these, absolute and relative (peak-continuum) OPD contributions are also calculated, with the peak-continuum OPD contribution calculated as the difference in the integrated PSD over the vibration peak width to the integrated local median over the same domain. Note this median detection method is efficient at detecting narrow band peaks but not optimized for wide band peaks. Therefore OPD peaks occuring on top of wide vibrations signals (such as is seen at 45Hz) are typically underestimated. Other detection methods were investigated but not reported here. In general we saw a diversity in the detected vibration peaks frequency and OPD contribution across different observations - highlighting the fact that the vibrational environment is complex and far from a static picture. As expected, detected vibration peaks in MN2 between 10-30Hz are generally much larger then those detected just before the Gravity fringe tracker since MN2 is optimized for this frequency band. Nevertheless, the contributions found before the Gravity fringe tracker in this frequency range are still comparable to dominant peaks found at higher frequencies. A series of dominant peaks can also typically be found in the 45-50Hz range for Gravity which are similarly matched in OPL level by the measured piston signal from the accelerometers - indicating that these peaks have significant contributions from the M1-M3, with figure \ref{fig:mn2_contr} providing evidence for which mirrors are the main culprits. However there are a series of higher frequency vibrations detected in the Gravity that are not detected in with the current accelerometers, this is common for all baselines across all dates - particularly in the 80-100Hz frequency range. Hence it is likely that these are caused downstream of the M3.  To further highlight the frequency domains that have the highest absolute OPD/OPL contribution, we integrate the Gravity pseudo open loop OPD and M1-M3 OPL PSDs for each respective baseline/telescopes in 10Hz bins, estimating the binned mean and standard deviation across all observations outlined in table \ref{tab:gravity_ft_obs_table}. These are plotted in figure \ref{fig:grav_v_mn2_freq_bi}. It is clear that the most damaging domains in Gravity's pseudo open loop are 10-20Hz, 40-50Hz, 80-100Hz, with the current accelerometers on the M1-M3 measuring a significant fraction of the power in the 10-20Hz and 40-50Hz bins, but not in the 80-100Hz range. Noting that the Gravity fringe tracker has greatest attenuation at lower frequencies, this result is in complete agreement with the initial analysis performed in section \ref{sec:vlti_vibrations} that vibration peaks around 80-90Hz are the most dominant contributor to the Gravity/UT OPD. Daytime metrology data suggest that these vibrations are coming from between the M4-M10 where the MN2 upgrade is planned. 

\section{ Conclusion }
This work looked at the theoretical OPD limitations of the VLTI and compared these to the measured values on the Gravity fringe tracker. The VLTI is capable of achieving sub 100nm RMS OPD. The Gravity OPD RMS on the UTs is roughly 42 times the theoretical atmospheric limit of $\sim$5nm RMS - which seems to be currently limited by vibrations along the VLTI optical train. Significant contributions are coming from vibrations in the 10-20Hz and 45-50Hz band that are currently seen on the Manhattan II system, in-addition to higher frequency contributions, particularly between 80-100Hz which are likely coming from between the M4-M10 mirrors. The Manhattan II upgrade will likely be capable of detecting these higher frequency contributions.  We also found the ATs to only be a factor of 6 above the theoretical limit of 15nm. Additionally, evidence was presented for the $f^{-17/3}$ power law in the OPD PSD of the AT's as expected theoretically from telescope filtering. This brings hope that there is significant room for improvement on the UT's. 

\appendix    

\acknowledgments 
A. Bigioli, D. Defrère and R. Laugier received funding from the European Research Council (ERC) under the European Union's Horizon 2020 research and innovation program (grant agreement CoG - 866070).

\bibliography{report} 
\bibliographystyle{spiebib} 

\end{document}